\documentclass[twocolumn,superscriptaddress,tightenlines,showpacs,floatfix,preprintnumbers, nofootinbib,hyperref]{revtex4} 
\pdfoutput=1
\usepackage[english]{babel}
\usepackage{amsmath,amssymb,amsfonts, bm,bbm,slashed, subdepth}
\usepackage{graphicx}
\usepackage[sort&compress]{natbib}
\usepackage{xcolor}
\usepackage[normalem]{ulem}
\usepackage{hyperref}
\usepackage{cleveref}
\definecolor{red}{rgb}{1.0, 0, 0}
\usepackage{hyperref}
\usepackage{enumerate}
\usepackage{epsfig, subfigure}
\usepackage{setspace}
\usepackage{booktabs, tabularx}
\usepackage{units}
\usepackage{hyperref}

\newcommand{\be}{\begin{equation}}
\newcommand{\ee}{\end{equation}}
\newcommand{\ba}{\begin{array}}
\newcommand{\ea}{\end{array}}
\newcommand{\bea}{\begin{eqnarray}}
\newcommand{\eea}{\end{eqnarray}}
\newcommand{\balg}{\begin{align}}
\newcommand{\ealg}{\end{align}}
\newcommand{\bit}{\begin{itemize}}
\newcommand{\eit}{\end{itemize}}
\newcommand{\trm}[1]{\textrm{#1}}

\newcommand{\Mpc}{\trm{\Mpc}}
\newcommand{\yr}{\trm{\yr}}
\newcommand{\eV}{\trm{\eV}}

\newcommand{\bd}[1]{{\color{black}#1}}
\newcommand{\zz}[1]{{\color{black}#1}}

\begin{document}

\title{Temporal Instability Enables Neutrino Flavor Conversions Deep Inside Supernovae}
\author{Basudeb Dasgupta}
\affiliation{Tata Institute of Fundamental Research,
             Homi Bhabha Road, Mumbai, 400005, India.}
\author{Alessandro Mirizzi}
\affiliation{Dipartimento Interateneo di Fisica ``Michelangelo Merlin,'' Via Amendola 173, 70126 Bari, Italy.\\
Istituto Nazionale di Fisica Nucleare - Sezione di Bari,
Via Amendola 173, 70126 Bari, Italy.}

\date{November 4, 2015}

\begin{abstract}
We show that a self-interacting neutrino gas can spontaneously acquire a non-stationary pulsating component in its flavor content, with a frequency that can exactly cancel the ``multi-angle'' refractive effects of dense matter. This can then enable homogeneous and inhomogeneous flavor conversion instabilities to exist \bd{even} at large neutrino and matter densities, where the system would have been stable if the evolution were strictly stationary. Large flavor conversions, \bd{especially} close to a supernova core, are possible via this novel mechanism. 
\bd{This may have} important consequences for the explosion dynamics, nucleosynthesis, as well as for neutrino observations of supernovae.

\end{abstract}


\maketitle

\section{Introduction} 
Inside a supernova (SN), neutrino densities are so high that neutrino flavor oscillations are affected not only by ordinary matter, but also by the neutrinos themselves~\cite{Pantaleone:1992xh,Pantaleone:1994ns,Qian:1994wh,Pastor:2002we}. Neutrino-neutrino interactions lead to highly correlated collective flavor conversions and unexpected effects, which completely change the physics of supernova neutrinos~\cite{Duan:2006an,Hannestad:2006nj,Fogli:2007bk,Fogli:2008pt,Dasgupta:2007ws,Dasgupta:2009mg,Dasgupta:2010ae,Friedland:2010sc,Dasgupta:2010cd,Cherry:2012zw,Raffelt:2013rqa,Mirizzi:2013rla,Mirizzi:2013wda,Duan:2013kba,Pehlivan:2011hp, Volpe:2013jgr, Vlasenko:2013fja, Serreau:2014cfa, Kartavtsev:2015eva}. See refs.\,\cite{Duan:2010bg,Mirizzi:2015eza} for recent reviews. \zz{In this work, we show that time-dependent fluctuations lead to a novel effect that can enable flavor conversion deeper in the SN than previously realized.}

Above the neutrinosphere, in the absence of collisions, the dynamics of a dense neutrino gas is characterized in terms of ``density matrices'' in flavor space, $\varrho(t,{\bf x},E,{\bf v})$ for neutrinos with energy $E$ and velocity ${\bf v}$ at position ${\bf x}$ and time $t$. These
obey the kinetic equations~\cite{Sigl:1992fn,Strack:2005ux,Vlasenko:2013fja},
\begin{eqnarray}
 i(\partial_t + {\bf v} \cdot \nabla)\varrho 
 = [{\sf H}, \varrho]
\,\ .
\label{eq:EiM}
\end{eqnarray}
On the left-hand-side (l.h.s.), the first term accounts for explicit time dependence, while the second, proportional to the neutrino velocity ${\bf v}$, is the drift-term due to neutrino free-streaming. On the right-hand-side (r.h.s.), ${\sf H}(t,{\bf x},E,{\bf v})$ is the Hamiltonian matrix in flavor space containing the {neutrino} mass-square matrix and potentials due to matter and neutrinos. 

{Flavor evolution of the dense neutrino gas, 
as 
governed by eq.\,(\ref{eq:EiM}), has a highly complex structure. It depends on the 4 time and space coordinates, the 4 energy and velocity coordinates (with $|{\bf v}|=1$, in our ultra-relativistic approximation), as well as the flavor states of all neutrinos. In order to reduce this complexity,} symmetries in the neutrino flavor evolution have often been assumed. For neutrinos in a SN environment, all of previous literature is based on the assumption that the evolution is stationary, i.e., there is no explicit time-dependence, or only a slow/small time-dependence 
that
does not significantly affect the flavor evolution. Additionally, under the assumption of a spherically symmetric neutrino emission, the dynamics reduces to a one-dimensional evolution along the radial coordinate. This is the rationale behind the often-used ``bulb model''~\cite{Duan:2006an,Fogli:2007bk}.

These symmetry assumptions,  viz., temporal stationarity and spatial homogeneity, have been recently criticized because self-interacting neutrinos can spontaneously break these space-time symmetries. Indeed, studies on simple toy-models show that the translation symmetries in time~\cite{Mangano:2014zda,Abbar:2015fwa} and space~\cite{Duan:2014gfa,Abbar:2015mca,Mirizzi:2015fva,Mirizzi:2015hwa,Chakraborty:2015tfa} are not stable. Even tiny 
\zz{inhomogeneities may lead to new flavor instabilities~\cite{Duan:2014gfa,Abbar:2015mca,Mirizzi:2015fva} that can develop also} 
\zz{at large neutrino densities, as above the SN core, where oscillations are otherwise expected to be suppressed due to synchronization.}
However, large neutrino densities in a SN are typically accompanied by a large matter density\,\cite{Chakraborty:2015tfa}, which produces ``multi-angle matter effects''~\cite{EstebanPretel:2008ni} that suppress both homogeneous and inhomogeneous instabilities. The current understanding is then that neutrinos cannot change their flavor too close to the SN core.

Flavor conversions at small distances from the SN core would have major consequences for SN explosions, nucleosynthesis, as well as neutrino observations of nearby SNe. If conversions are possible below the shock radius, neutrinos can provide a net positive energy to the shock and assist SN explosions~\cite{fullershock,Suwa:2011ac, Dasgupta:2011jf, Pejcha:2011en}. Similarly, the neutron-to-proton ratio can be changed deeper inside a star, affecting the yield of heavier nuclei created through the r-process~\cite{Duan:2010af,Pllumbi:2014saa}. Also, in order to interpret any potential observation of neutrinos from SNe, current and proposed neutrino experiments depend crucially on understanding where and how the flavor-dependent neutrino fluxes have converted to each other~\cite{Dasgupta:2011wg,Serpico:2011ir,Borriello:2012zc,Mirizzi:2015eza}. Now that Gd-doping in Super-Kamiokande~\cite{Beacom:2003nk} is approved~\cite{SuperK-Gd}, the imminent observation of the diffuse background of SN neutrinos may raise this issue~\cite{Lunardini:2012ne}, \zz{even without a Galactic SN.}

In the following, using linear stability analysis we show the presence of \zz{an unstable pulsating mode} that leads to flavor conversion at high neutrino and matter density. The key insight is that the frequency of pulsation can undo the phase dispersion due to a large matter density. As a result, flavor instabilities, which would have grown only if matter effects were small, can now develop at large neutrino and matter densities.
Then, to demonstrate that this linear instability survives in the non-linear regime, we numerically calculate the flavor evolution in a simplified model
and show that flavor conversions indeed occur at large neutrino and matter densities when there are space and time-dependent fluctuations. Finally, we discuss the implications for SN neutrinos and conclude.

\section{Linear analysis for a general scenario}
Assuming that the neutrinos are initially in flavor eigenstates, their density matrices $\varrho(t,{\bf x},E,{\bf v})$ can be written {in} a 2-flavor framework as
\begin{align}
\varrho=\frac{{\rm Tr}(\varrho)}{2}
+\frac{n_\nu}{2}g
\begin{pmatrix}
1 & S\\ 
 S^*&-1 
\end{pmatrix}\,,
\label{eq:lin}
\end{align}
to linear order in $S(t,{\bf x},E, {\bf v})$~\cite{Banerjee:2011fj}. The quantity $g({t, {\bf x},E,{\bf v}})$ is the energy and angular distribution of neutrinos from the source and $n_\nu$ is an arbitrary normalization constant, with dimensions of number density, for making $S$ dimensionless. {A non-zero} off-diagonal element $S$ {represents flavor conversions}. For antineutrinos, $\bar\varrho(E)\equiv-\varrho(-E)$, {extending} the physical range of $E$ from $-\infty$ to $+\infty$. The Hamiltonian for the flavor evolution is
\begin{align}
{\sf H}=\frac{\sf M^2}{2E}+\sqrt{2}G_F{\sf N}_l+ \sqrt{2}G_F\int\,d\Gamma'(1-{\bf v}\cdot{\bf v}')\varrho'\,,
\label{eq:ham}
\end{align}
where sans-serif quantities are $2\times2$ matrices in flavor space. {Namely, ${\sf M^2}$ is the neutrino mass-squared matrix, while $\sqrt{2}G_F{\sf N}_l$ and the last term on the r.h.s. appear due to forward scattering on matter~\cite{Matt} and neutrinos~\cite{Pantaleone:1992xh,Pantaleone:1994ns}, respectively. The integral is over all neutrino {energies and velocities}, i.e., $\int d\Gamma'=\int_{-\infty}^{+\infty}dE'\,E'^2\int d{\bf v}'/(2\pi)^3$.}

In a non-isotropic neutrino gas, as in the case of neutrinos streaming off a SN core, there is a net neutrino current so that neutrinos moving in different directions, i.e., with different ${\bf v}$, acquire different phases via the velocity-dependent terms, i.e., $(1-{\bf v}\cdot{\bf v}')$ {in ${\sf H}$} and 
${\bf v}\cdot\nabla$ from the {drift-term} in eq.~(\ref{eq:EiM}). These are multi-angle effects, that arise due to the current-current nature of the low-energy weak interactions and the source geometry. 
\zz{Typically, they inhibit the collective behavior of the flavor evolution, but can also lead to flavor decoherence~\cite{Raffelt:2007yz,EstebanPretel:2007ec,Sawyer:2008zs}.}

Using eqs.\,(\ref{eq:lin}) and (\ref{eq:ham}) in eq.\,(\ref{eq:EiM}), and taking a vanishing mixing angle, we find the equation for flavor evolution,
\begin{align}
i(\partial_t + {\bf v} \cdot \nabla)S=\big[-\omega+\lambda&+\mu\mbox{$\int$}\,d\Gamma'(1-{\bf v}\cdot{\bf v}')g'\big]S \nonumber \\
&-\mu\mbox{$\int$}\,d\Gamma'(1-{\bf v}\cdot{\bf v}')g'S'\,,
\label{eq:master}
\end{align}
where the relevant energy scales are the neutrino oscillation frequency in vacuum $\omega=\Delta m^2/(2E)$, the matter potential $\lambda=\sqrt{2}G_F n_e$, and the neutrino potential $\mu=\sqrt{2}G_F\,n_\nu$. Note that $\mu$ always appears in product with $S$, making the precise choice of $n_\nu$ immaterial.

Let us consider the evolution of $S$ {along the radial distance $r$, while Fourier decomposing it in $t$ and the spatial coordinates transverse to $\hat{r}$, viz., ${\bf r}_T$.}  We will take the spectrum $g'({t, {\bf x},E,{\bf v}})$ to be independent of time and space, i.e.,
$g'({t, {\bf x},E,{\bf v}})\equiv g'(E,{\bf v})$, so that it does not get Fourier transformed. Explicitly,
\begin{align}
S=\int_{-\infty}^{+\infty}d p\,d{\bf k}\,e^{-i( pt + {\bf k}\cdot{\bf r}_T)}Q_{p,{\bf k}}e^{-i\Omega_{\zz{p,{\bf k}}}\,r}\,,
\label{eq:ft}
\end{align}
where $Q_{p,{\bf k}}e^{-i\Omega_{\zz{p,{\bf k}}}\,r}$ is the Fourier coefficient of a flavor evolution mode with temporal pulsation $p$ and {inhomogeneity} wavevector ${\bf k}$ . Inserting this ansatz into eq.\,(\ref{eq:master}), using ${\bf v}\cdot \nabla=v_r\partial_r+{\bf v}_T\cdot\nabla_T$, and dividing by the radial velocity $v_r$, we  find an eigenvalue equation for $Q_{p,{\bf k}}(E, {\bf v})$,
\begin{align}
\bigg[\frac{-\omega+\bar\lambda-p-{\bf v}_T\cdot{\bf k}}{v_r}\,-\,&\Omega_{\zz{p,{\bf k}}}\bigg]Q_{p,{\bf k}}=\nonumber\\
\frac{\mu}{v_r}\int\,d\Gamma'&(1-{\bf v}\cdot{\bf v}')g'Q'_{p,{\bf k}}\,,
\label{eq:eigen}
\end{align}
where $\bar\lambda=\lambda+\mu\int\,d\Gamma'(1-{\bf v}\cdot{\bf v}')g'$ encodes ``matter" effects from both matter and neutrinos. Note that in the linear regime, different Fourier modes are not coupled. A growing solution to this equation, with ${\rm Im}(\Omega)>0$, signals that there is an instability.

For a stationary system one finds 
growing solutions even at a large neutrino density, if inhomogeneities are present, i.e., ${\bf k}\neq0$, as long as $\bar\lambda\ll\mu$~\cite{Duan:2014gfa,Chakraborty:2015tfa,Abbar:2015mca}.  
\zz{In a SN however, $\bar\lambda$ is also large when $\mu$ is large and these instabilities are typically not realized~\cite{Chakraborty:2015tfa}.}
The reason for this is clear: One cannot simultaneously 
\bd{obtain growing solutions}
for all velocities, because the inhomogeneous term, the matter term, and the $\mu$-dependent neutrino-neutrino interaction term on the r.h.s. are all large, i.e., ${\bf v}_T\cdot{\bf k},\,\bar\lambda,\,\mu\gg\omega$, but have different velocity-dependences which cannot completely cancel against each other.

The non-stationary system has an innocuous-looking but important difference with respect to the stationary system. Non-stationarity lowers $\bar\lambda$ by $p$, i.e., $\bar\lambda\rightarrow \bar\lambda-p$, 
\bd{as also seen in ref.\,\cite{Abbar:2015fwa}.}
\bd{More importantly, a fact not realized so far}, this $p\neq0$ term has the same \zz{multi-angle} dependence as $\bar\lambda$. Therefore, if one allows for a non-stationary solution, the neutrino system with a pulsation $p\simeq\bar\lambda$ can undo the phase dispersion due to a large matter term for all velocities. Thus, one can find growing solutions, with ${\rm Im}(\Omega)>0$, to the eigenvalue equation [eq.\,(\ref{eq:eigen})], as previously for the $\bar\lambda\ll\mu$ scenario.  \bd{The eigenvalues are identical to those in Sec.\,4.2.4 of ref.\,\cite{Chakraborty:2015tfa} with the shift $\bar\lambda\rightarrow \bar\lambda-p$.} These solutions are highly oscillatory in space (${\bf k}\neq0$) and time ($p\neq0$), and would lead to flavor averaging. \emph{This is our main result.}

\begin{figure}[!t]
  \centering
  \includegraphics[width=0.85\columnwidth]{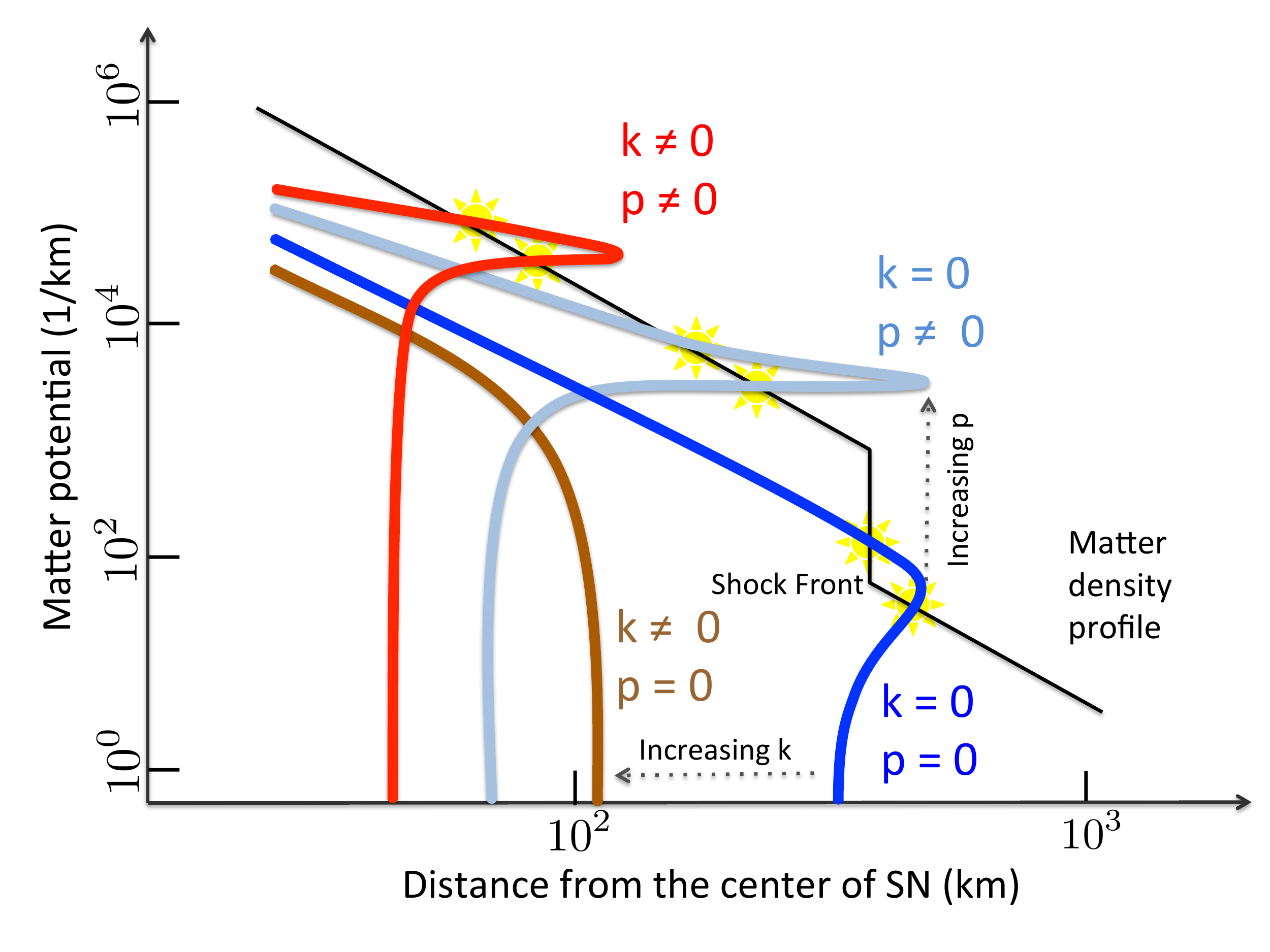}
  \caption{Schematic of the SN matter potential (thin line) and densities where neutrino flavor composition is unstable to small oscillations with spatial inhomogeneity wavenumber $k$ and frequency of time-dependence $p$ (thick lines)~\cite{pvtcom}. Time-dependent fluctuations can be unstable at higher matter density and lead to neutrino flavor conversions deeper in a SN.}
  \label{fig:1}
\end{figure}
\zz{In Fig.\,\ref{fig:1}, we illustrate this general idea in the context of SN neutrinos. The thin black line shows a SN density profile, while the thick colored lines schematically show where flavor instabilities with certain pulsation $p$ and wavenumber $k=|{\bf k}|$ can grow, i.e., have ${\rm Im}(\Omega)>0$. The inhomogeneous instability, i.e., with ${k}\neq0$, is always \emph{below} the homogeneous ${k}=0$ instability~\cite{Chakraborty:2015tfa}, and never occurs for the physically available matter density. However, $p\simeq\bar\lambda$ can \emph{raise} these ${k}=0$ and ${k}\neq0$ instabilities, making them unstable at low radii. They can now develop at large $\mu$ and $\lambda$, i.e., at a small radius, with a temporal oscillation of frequency $\simeq p$. This can happen for both normal and inverted neutrino mass ordering.}

How does this linear instability evolve when linear theory is no longer appropriate? What is its impact? To answer these questions, we must numerically solve the equations of motion (EoMs), eq.\,(\ref{eq:EiM}), in the fully non-linear regime. This is what we do next.

\section{Non-linear analysis for the  two-beams model}
The new effect, relevant to SN neutrinos, is that a stationary system, which is stable to both homogeneous and inhomogeneous perturbations at large $\mu$ and $\lambda$, becomes unstable therein when non-stationarity is allowed. This requires simulating a system with temporal non-stationarity, spatial inhomogeneity, and multi-angle matter effects, which is extremely challenging and has not been attempted so far. Here, we present the first simulation with these three features.

The simplest model that can accommodate the required features is the neutrino ``line model''~\cite{Duan:2014gfa,Mirizzi:2015fva,Chakraborty:2015tfa}. In this model, one considers  monochromatic  neutrinos emitted in two directions, ``$L$\,'' and ``$R$\,'', from an infinite plane at $z=0$. Assuming translational invariance along the $y$-direction, the flavor evolution along  $z >0$ can be characterized  on the two-dimensional plane  spanned by the $x$ and $z$ coordinates. The neutrino emission modes $L$ and $R$ are labeled  in terms of their velocities, i.e., $ { {\bf v}}_{L} = (v_{x,L}, 0,v_{z,L})=(\cos \vartheta _{L},0,\sin \vartheta _{L})$, where $\vartheta_{L} \in [0, \pi]$ is the emission angle, and similarly for ${ {\bf v}}_{R}$. Thus, for the $L$ mode, the differential operator on the l.h.s. in eq.\,(\ref{eq:EiM}) takes the form 
\begin{equation}
 \partial_t + { {\bf v}}_{L} \cdot \nabla= \partial_t + v_{x,L} \partial_x + v_{z,L} \partial_z\,, 
 \label{eq:diffoper}
\end{equation}
while the  Hamiltonian in eq.\,(\ref{eq:ham}) becomes
\begin{align}
{\sf H}_L=\frac{-\omega+\lambda}{2}{\sigma}_3+ \mu(1-{\bf v}_L\cdot{\bf v}_R) \left[(1+\epsilon)\varrho_R -\bar\varrho_R\right]\,,
\label{eq:hamL}
\end{align}
where $\sigma_3$ is the diagonal Pauli matrix, and $\epsilon$ is the neutrino-antineutrino asymmetry, i.e., $1+\epsilon=(n_{\nu_e}-n_{\nu_x})/(n_{\bar\nu_e}-n_{\bar\nu_x})$, with $\nu_x$ being a non-electron flavor. The normalization $n_\nu=n_{\bar\nu_e}-n_{\bar\nu_x}$ is used to define $\mu=\sqrt{2}G_F\,(n_{\bar\nu_e}-n_{\bar\nu_x})$.

The differential operator in eq.\,(\ref{eq:diffoper}) shows that the flavor evolution is determined by a partial differential equation in one temporal and two spatial dimensions. By Fourier transforming the EoMs in $t$ and $x$, as in eq.\,(\ref{eq:ft}), one obtains a tower of ordinary differential equations in the $z$-coordinate for the different Fourier modes $\varrho_{p,k}$ with temporal pulsation $p$ and spatial wavenumber ${k}$. In the non-linear regime, the EoMs for the different Fourier modes have a convolution term due to interactions between the different modes~\cite{Mirizzi:2015fva}.

If $v_{z,L}=v_{z,R}$, as assumed in refs.\,\cite{Duan:2014gfa,Mirizzi:2015fva}, even a very large matter term $\lambda$ can be rotated away from the EoMs by  studying the flavor evolution in a suitable co-rotating frame~\cite{Hannestad:2006nj}. 
\zz{Conversely, if $v_{z,L} \neq v_{z,R}$ the matter term leads to frequencies $\lambda/{v_{z,L}}$ for the $L$ mode and $\lambda/{v_{z,R}}$ for the $R$ mode. Their difference cannot be removed and gives multi-angle matter effects~\cite{EstebanPretel:2008ni}.}
If $\lambda \gg \mu$, \zz{this} phase difference between $L$ and $R$ modes is so large that it suppresses the self-induced flavor conversions from both $k=0$ and $k\neq0$ instabilities~\cite{EstebanPretel:2008ni,Chakraborty:2015tfa}. Conversely, if one allows for a non-stationary solution, the neutrino system selects the pulsations $p\simeq\bar\lambda$, that compensate the phase dispersion due to a large matter term, and generate growing instabilities, in particular at small-scales associated with spatial inhomogeneities.

To quantitatively illustrate this claim, we take the source at $z=0$ to emit only $\nu_e$ and $\bar\nu_e$, with an factor of two excess of $\nu_e$ over $\bar\nu_e$, i.e., $\epsilon=1$. We choose $\theta=10^{-3}$ and a normal mass ordering, i.e., $\omega>0$, but the result would be similar for the inverted ordering, i.e., $\omega<0$. The overall frequency-scale is set by $\omega=1$. A large $\mu=40$ is chosen, so that oscillations are suppressed in the homogeneous case, as in a SN.  We take the $L$ and $R$ modes to have two different angles $\vartheta_R= 5 \pi/18$ and $\vartheta_L=  7\pi/9$, so that a large matter \zz{potential}, $\lambda=4\times 10^4$, suppresses the inhomogeneous modes, mimicking the similar effect in a SN. 


\begin{figure*}[!t]
  \includegraphics[width=0.8\textwidth]{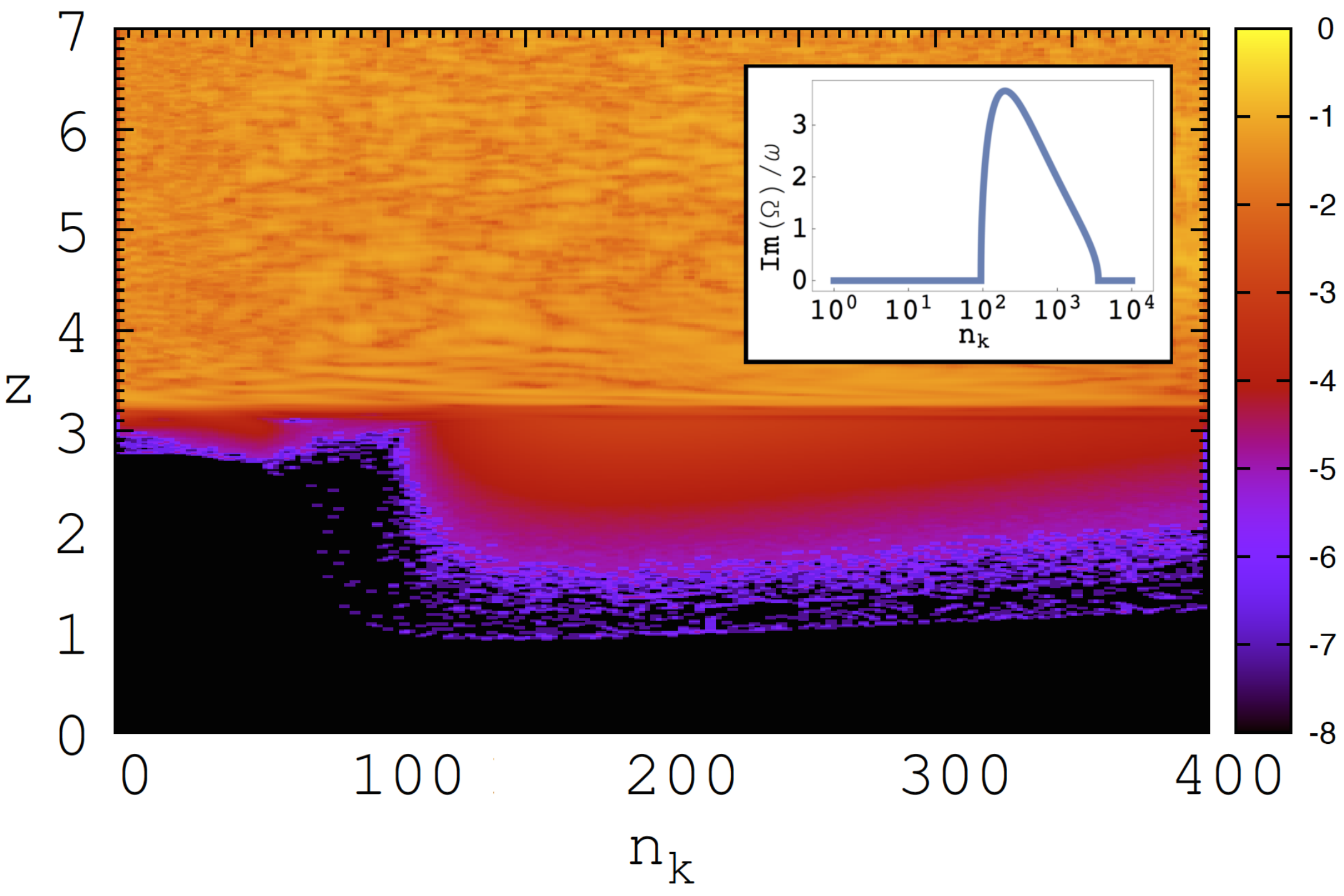}
  \caption{\zz{Amplitudes of flavor conversion at distance $z$ in our two-beams model, ${\rm log}_{10}\big|\varrho^{e\mu}_{n_k}(z)|$, for the $n_k$ Fourier modes of inhomogeneity along the transverse direction. Time-dependent pulsations of frequency $p=\bar\lambda$ and large neutrino and matter densities are included. Inset: Linear growth rates ${\rm Im}(\Omega)/\omega$ for a larger range of $n_k$ (note the log-scale). Modes $n_k\simeq10^2$ grow first and quickly excite all other modes, leading to large flavor conversion.}}
  \label{fig:2}
\end{figure*}

In order to make $k$ and $p$ dimensionless, $k$ is expressed in multiples of the vacuum oscillation frequency, i.e., $k= n_k\,\omega$, while $p$ is expressed in multiples of the matter potential $\bar\lambda$, i.e., $p = n_p\,\bar\lambda$. For simplicity, we limit ourselves to the $n_p=1$ mode. In this way, in the non-linear regime we only have to consider the convolution among the different Fourier modes associated with spatial inhomogeneities,
\zz{$\sim \sum_{j_k}[((1+\epsilon)\varrho_{R, n_k-j_k} -\bar\varrho_{R,n_k-j_k}), \varrho_{L,j_k}]$}, 
and analogously for the $R$ mode~\cite{Mirizzi:2015fva}. \zz{We include modes up to $n_k=600$ but ensure that $n_k > 400$ remain empty. This trick avoids ``spectral blocking'' that leads to a spurious rise of the Fourier coefficients at large $n_k$ due to truncation of the tower of equations~\cite{spectral}.} To seed the spatial inhomogeneity, we use numerical noise of ${{\mathcal O}(10^{-8})}$ for all modes.

\zz{Fig.\,\ref{fig:2} shows the flavor dynamics of this two-beams model. Amplitudes of flavor conversion, ${\rm log}_{10}\big|\varrho^{e\mu}_{n_k}(z)|$, are shown at distance $z$ for various Fourier modes $n_k$ of inhomogeneity along $x$, in the presence of time-dependent fluctuations with frequency $p=\bar\lambda$.  As linear theory predicts [${\rm Im}(\Omega)$~vs.\,$n_k$ shown in the inset], modes around $n_k\simeq100$ are the most unstable and grow first with the predicted rate ${\rm Im}(\Omega)/\omega\simeq3$. However, while only the modes $n_k\simeq{\cal O}(10^2)$ are unstable according to linear analysis, a cascade in Fourier space develops to both smaller and larger $n_k$ due to the convolution enforced by the neutrino-neutrino interaction. Modes with $n_k<10^2$ also grow fast when nonlinearity sets in. The flavor composition begins to oscillate with many frequencies and appears to be ``averaged out''.}

Thus, we find that neutrinos can change flavor at large $\lambda$ and $\mu$, if non-stationary solutions with frequency $p\simeq\lambda$ are allowed. We expect that the cascade in Fourier space leads to ``flavor decoherence'', or approximate equilibration between all flavors~\cite{Sawyer:2005jk}. In a SN, this may have important consequences, which we discuss below. 
  
\section{Discussion and conclusions}  
An important question is if this effect can be important in a SN. Here, we provide a back-of-the-envelope estimate. \zz{To aid shock revival,} flavor instability has to occur below the shock-front at $r\simeq200\,$km and above the gain radius at $r\simeq100\,$km (see, e.g., Fig.\,4 in ref.\,\cite{Dasgupta:2011jf}). One thus needs pulsations of high frequency $p\simeq\lambda\sim10^{3-6}\,$km$^{-1}\sim3\times 10^{8-11}\,{\rm Hz}$. Between $r\simeq100$\,km to $150$\,km, a typical instability with growth rate ${\rm Im}(\Omega)/\omega\simeq 3$ then grows by $\sim60$ e-foldings, i.e., a factor of $\simeq10^{26}$, for $15\,$MeV neutrinos with $\omega\sim 0.4~{\rm km}^{-1}$, \bd{assuming constant growth}. Can such high-frequency fluctuations occur in a SN with even tiny amplitudes? In this context, we find it intriguing that pair-correlations of the neutrino field, which are many-body corrections to the single-particle density matrices, lead to relative number fluctuations of a size $\kappa^2\sim (\lambda \beta/E)^2\sim10^{-22}$, where $\beta\simeq10^{-2}c$ is the typical speed of ordinary matter in SN, which oscillate with a frequency $\sim 2E\sim10^{22}{\rm Hz}$ for $15\,$MeV neutrinos, as shown in ref.~\cite{Kartavtsev:2015eva}.
\zz{However, in a realistic SN the density behind the shock, though often much flatter than shown in Fig.\,\ref{fig:1}, is not \emph{exactly} constant and the growth rates decrease when $p\not\simeq\bar\lambda$. On the other hand, then nearby $p$ modes are excited and nonlinear coupling of modes makes all instabilities grow. More detailed studies are needed to study these effects~\cite{wip}, the impact on nucleosynthesis, and overall flavor conversion.}

Potentially, the consequences of our finding augur another paradigm-shift in the understanding of self-induced conversions and on their impact on the SN dynamics. The possibility of low-radii conversions behind the stalled shock-wave during the accretion phase, suppressed by the large matter term in the stationary and homogeneous case~\cite{Chakraborty:2011nf,Chakraborty:2011gd,Sarikas:2011am,Chakraborty:2014nma,Chakraborty:2014lsa}, implies that the flavor dynamics may need to be taken into account in the revitalization of the shock-wave~\cite{fullershock,Dasgupta:2011jf}. Also, the impact on nucleosynthesis in a SN would be important~\cite{Duan:2010af,Pllumbi:2014saa}. With flavor equilibration, the interpretation of observed SN fluxes may also become simpler~\cite{Sawyer:2005jk}.

However, the possibility of  flavor conversions  close  to  the neutrinosphere in a SN  also questions the assumption that flavor conversions safely occur outside it. This assumption allowed one to replace the full Boltzmann equations, containing both oscillations and scatterings, with the flavor oscillation equations for free-streaming neutrinos. In a non-stationary situation, this assumption may no longer be guaranteed and imply the necessity to simultaneously perform the neutrino transport and flavor evolution~\cite{Vlasenko:2013fja}. This is a formidable problem that would require new computational techniques.  Furthermore, it is possible that these instabilities (inhomogeneity and non-stationarity) appear in a regime where the coarse-grained description adopted using density matrices~\cite{Pantaleone:1994ns} is insufficient. Although we have used this standard description here, as a first step, this is a more fundamental aspect that needs further studies.

In conclusion, we have presented the first study of nonlinear effects of non-stationarity in a dense neutrino gas. We have pointed out novel temporal instabilities that can dramatically affect flavor evolution, and raise the possibility of self-induced flavor conversions deep in a SN. The discovery of the role of symmetry-breaking in the flavor evolution of SN neutrinos is opening completely new directions of investigations. We foresee that many surprises are still in store.

\section{Acknowledgments}
We warmly thank A.\,Dighe,  S.\,Shalgar, and H.\,Vogel for useful discussions, and especially \zz{S.\,Chakraborty, I.\,Izaguirre,} E.\,Lisi, G.\,Raffelt, and A.\,Smirnov for a careful reading of the manuscript and useful comments to improve it. B.D. thanks the organizers of TAUP 2015 at Torino, whose invitation motivated this work. The work of A.M. is supported by the Italian Ministero dell'Istruzione, Universit\`a e Ricerca (MIUR) and Istituto Nazionale di Fisica Nucleare (INFN) through the ``Theoretical Astroparticle Physics'' project.


\begin{thebibliography}{99}


\bibitem{Pantaleone:1992xh} 
  J.~T.~Pantaleone,
  ``Dirac neutrinos in dense matter,''
  Phys.\ Rev.\ D {\bf 46}, 510 (1992).

\bibitem{Pantaleone:1994ns} 
  J.~T.~Pantaleone,
  ``Neutrino flavor evolution near a supernova's core,''
  Phys.\ Lett.\ B {\bf 342}, 250 (1995)
  [astro-ph/9405008].
  
\bibitem{Qian:1994wh} 
  Y.~Z.~Qian and G.~M.~Fuller,
  ``Neutrino-neutrino scattering and matter enhanced neutrino flavor transformation in Supernovae,''
  Phys.\ Rev.\ D {\bf 51}, 1479 (1995)
  [astro-ph/9406073].
  
\bibitem{Pastor:2002we}
  S.~Pastor and G.~Raffelt,
  ``Flavor oscillations in the supernova hot bubble region: Nonlinear effects of neutrino background,''
  Phys.\ Rev.\ Lett.\  {\bf 89} (2002) 191101
  [astro-ph/0207281].
  
 
\bibitem{Duan:2006an} 
  H.~Duan, G.~M.~Fuller, J.~Carlson and Y.~-Z.~Qian,
  ``Simulation of Coherent Non-Linear Neutrino Flavor Transformation in the Supernova Environment. 1. Correlated Neutrino Trajectories,''
  Phys.\ Rev.\ D {\bf 74}, 105014 (2006)
  [astro-ph/0606616].
 

\bibitem{Hannestad:2006nj}
  S.~Hannestad, G.~G.~Raffelt, G.~Sigl and Y.~Y.~Y.~Wong,
  ``Self-induced conversion in dense neutrino gases: Pendulum in flavour  
 space,''
  Phys.\ Rev.\  D {\bf 74}, 105010  (2006)
  [Erratum-ibid.\  D {\bf 76},  029901 (2007)]
  [astro-ph/0608695].

 
\bibitem{Fogli:2007bk} 
  G.~L.~Fogli, E.~Lisi, A.~Marrone and A.~Mirizzi,
  ``Collective neutrino flavor transitions in supernovae and the role of trajectory averaging,''
  JCAP {\bf 0712}, 010 (2007)
  [arXiv:0707.1998 [hep-ph]].
  
  
\bibitem{Fogli:2008pt} 
  G.~L.~Fogli, E.~Lisi, A.~Marrone, A.~Mirizzi and I.~Tamborra,
  ``Low-energy spectral features of supernova (anti)neutrinos in inverted hierarchy,''
  Phys.\ Rev.\ D {\bf 78}, 097301 (2008)
  [arXiv:0808.0807 [hep-ph]].
  
\bibitem{Dasgupta:2007ws} 
  B.~Dasgupta and A.~Dighe,
  ``Collective three-flavor oscillations of supernova neutrinos,''
  Phys.\ Rev.\ D {\bf 77}, 113002 (2008)
  [arXiv:0712.3798 [hep-ph]].
  
\bibitem{Dasgupta:2009mg}
  B.~Dasgupta, A.~Dighe, G.~G.~Raffelt and A.~Y.~Smirnov,
  ``Multiple Spectral Splits of Supernova Neutrinos,''
  Phys.\ Rev.\ Lett.\  {\bf 103} (2009) 051105
  [arXiv:0904.3542 [hep-ph]].
  
   
\bibitem{Friedland:2010sc} 
  A.~Friedland,
  ``Self-refraction of supernova neutrinos: mixed spectra and three-flavor instabilities,''
  Phys.\ Rev.\ Lett.\  {\bf 104}, 191102 (2010)
  [arXiv:1001.0996 [hep-ph]].
  
\bibitem{Dasgupta:2010ae}
  B.~Dasgupta, G.~G.~Raffelt and I.~Tamborra,
  ``Triggering collective oscillations by three-flavor effects,''
  Phys.\ Rev.\ D {\bf 81} (2010) 073004
  [arXiv:1001.5396 [hep-ph]].

  
\bibitem{Dasgupta:2010cd} 
  B.~Dasgupta, A.~Mirizzi, I.~Tamborra and R.~Tom{\`a}s,
  ``Neutrino mass hierarchy and three-flavor spectral splits of supernova neutrinos,''
  Phys.\ Rev.\ D {\bf 81}, 093008 (2010)
  [arXiv:1002.2943 [hep-ph]].
  
  
\bibitem{Cherry:2012zw} 
  J.~F.~Cherry, J.~Carlson, A.~Friedland, G.~M.~Fuller and A.~Vlasenko,
  ``Neutrino scattering and flavor transformation in supernovae,''
  Phys.\ Rev.\ Lett.\  {\bf 108}, 261104 (2012)
  [arXiv:1203.1607 [hep-ph]].
  
\bibitem{Raffelt:2013rqa} 
  G.~Raffelt, S.~Sarikas and D.~de Sousa Seixas,
  ``Axial Symmetry Breaking in Self-Induced Flavor Conversion of Supernova Neutrino Fluxes,''
  Phys.\ Rev.\ Lett.\  {\bf 111}, no. 9, 091101 (2013)
  [Phys.\ Rev.\ Lett.\  {\bf 113}, no. 23, 239903 (2014)]
  [arXiv:1305.7140 [hep-ph]].
  
\bibitem{Mirizzi:2013rla} 
  A.~Mirizzi,
  ``Multi-azimuthal-angle effects in self-induced supernova neutrino flavor conversions without axial symmetry,''
  Phys.\ Rev.\ D {\bf 88}, no. 7, 073004 (2013)
  [arXiv:1308.1402 [hep-ph]].
  
\bibitem{Mirizzi:2013wda} 
  S.~Chakraborty and A.~Mirizzi,
  ``Multi-azimuthal-angle instability for different supernova neutrino fluxes,''
  Phys.\ Rev.\ D {\bf 90}, no. 3, 033004 (2014)
  [arXiv:1308.5255 [hep-ph]].
  
\bibitem{Duan:2013kba} 
  H.~Duan,
  ``Flavor Oscillation Modes In Dense Neutrino Media,''
  Phys.\ Rev.\ D {\bf 88}, 125008 (2013)
  [arXiv:1309.7377 [hep-ph]].
  
\bibitem{Pehlivan:2011hp}
  Y.~Pehlivan, A.~B.~Balantekin, T.~Kajino and T.~Yoshida,
  ``Invariants of Collective Neutrino Oscillations,''
  Phys.\ Rev.\ D {\bf 84} (2011) 065008
  [arXiv:1105.1182 [astro-ph.CO]].
  
\bibitem{Volpe:2013jgr}
  C.~Volpe, D.~ V\"a\"an\"anen and C.~Espinoza,
  ``Extended evolution equations for neutrino propagation in astrophysical and cosmological environments,''
  Phys.\ Rev.\ D {\bf 87} (2013) 11,  113010
  [arXiv:1302.2374 [hep-ph]].

\bibitem{Vlasenko:2013fja} 
  A.~Vlasenko, G.~M.~Fuller and V.~Cirigliano,
  ``Neutrino Quantum Kinetics,''
  Phys.\ Rev.\ D {\bf 89}, no. 10, 105004 (2014)
  [arXiv:1309.2628 [hep-ph]].
  
\bibitem{Serreau:2014cfa} 
  J.~Serreau and C.~Volpe,
  ``Neutrino-antineutrino correlations in dense anisotropic media,''
  Phys.\ Rev.\ D {\bf 90}, no. 12, 125040 (2014)
  [arXiv:1409.3591 [hep-ph]].
  
\bibitem{Kartavtsev:2015eva}
  A.~Kartavtsev, G.~Raffelt and H.~Vogel,
  ``Neutrino propagation in media: Flavor-, helicity-, and pair correlations,''
  Phys.\ Rev.\ D {\bf 91} (2015) 12,  125020
  [arXiv:1504.03230 [hep-ph]].
  
\bibitem{Duan:2010bg} 
  H.~Duan, G.~M.~Fuller and Y.~-Z.~Qian,
  ``Collective Neutrino Oscillations,''
  Ann.\ Rev.\ Nucl.\ Part.\ Sci.\  {\bf 60}, 569 (2010)
  [arXiv:1001.2799 [hep-ph]].
  
\bibitem{Mirizzi:2015eza} 
  A.~Mirizzi, I.~Tamborra, H.~T.~Janka, N.~Saviano, K.~Scholberg, R.~Bollig, L.~Hudepohl and S.~Chakraborty,
  ``Supernova Neutrinos: Production, Oscillations and Detection,''
  arXiv:1508.00785 [astro-ph.HE].
  

\bibitem{Sigl:1992fn}  
  G.~Sigl and G.~Raffelt,  
  ``General kinetic description of relativistic mixed neutrinos,''  
  Nucl.\ Phys.\ B {\bf 406}, 423 (1993).  
  
\bibitem{Strack:2005ux} 
  P.~Strack and A.~Burrows,
  ``Generalized Boltzmann formalism for oscillating neutrinos,''
  Phys.\ Rev.\ D {\bf 71}, 093004 (2005)
  [hep-ph/0504035].
  

  
\bibitem{Mangano:2014zda} 
  G.~Mangano, A.~Mirizzi and N.~Saviano,
  ``Damping the neutrino flavor pendulum by breaking homogeneity,''
  Phys.\ Rev.\ D {\bf 89}, no. 7, 073017 (2014)
  [arXiv:1403.1892 [hep-ph]].
  
\bibitem{Abbar:2015fwa} 
  S.~Abbar and H.~Duan,
  ``Neutrino flavor instabilities in a time-dependent supernova model,''
  arXiv:1509.01538 [astro-ph.HE].

\bibitem{Duan:2014gfa} 
  H.~Duan and S.~Shalgar,
  ``Flavor instabilities in the neutrino line model,''
  Phys.\ Lett.\ B {\bf 747}, 139 (2015)
  [arXiv:1412.7097 [hep-ph]].

\bibitem{Abbar:2015mca} 
  S.~Abbar, H.~Duan and S.~Shalgar,
  ``Flavor instabilities in the multi-angle neutrino line model,''
  arXiv:1507.08992 [hep-ph].

\bibitem{Mirizzi:2015fva} 
  A.~Mirizzi, G.~Mangano and N.~Saviano,
  ``Self-induced flavor instabilities of a dense neutrino stream in a two-dimensional model,''
  Phys.\ Rev.\ D {\bf 92}, no. 2, 021702 (2015)
  [arXiv:1503.03485 [hep-ph]].

\bibitem{Mirizzi:2015hwa} 
  A.~Mirizzi,
  ``Breaking the symmetries of the bulb model in two-dimensional self-induced supernova neutrino flavor conversions,''
  arXiv:1506.06805 [hep-ph].

\bibitem{Chakraborty:2015tfa} 
  S.~Chakraborty, R.~S.~Hansen, I.~Izaguirre and G.~Raffelt,
  ``Self-induced flavor conversion of supernova neutrinos on small scales,''
  arXiv:1507.07569 [hep-ph].


\bibitem{EstebanPretel:2008ni} 
  A.~Esteban-Pretel, A.~Mirizzi, S.~Pastor, R.~Tom{\`a}s, G.~G.~Raffelt, P.~D.~Serpico and G.~Sigl,
  ``Role of dense matter in collective supernova neutrino transformations,''
  Phys.\ Rev.\ D {\bf 78}, 085012 (2008)
  [arXiv:0807.0659 [astro-ph]].
  
 \bibitem{fullershock}
G.~M.~Fuller, R.~Mayle, S.~Bradley Meyer, J.R.~Wilson, 
  ``Can a closure mass neutrino help solve the supernova shock reheating problem?,''
  Astrophys.\ Journ.\ Part 1, vol.~389, p.~517-526 (1992).
  
  
\bibitem{Suwa:2011ac}
  Y.~Suwa, K.~Kotake, T.~Takiwaki, M.~Liebendorfer and K.~Sato,
  ``Impacts of collective neutrino oscillations on core-collapse supernova explosions,''
  Astrophys.\ J.\  {\bf 738} (2011) 165
  [arXiv:1106.5487 [astro-ph.HE]].
  
\bibitem{Dasgupta:2011jf} 
  B.~Dasgupta, E.~P.~O'Connor and C.~D.~Ott,
  ``The Role of Collective Neutrino Flavor Oscillations in Core-Collapse Supernova Shock Revival,''
  Phys.\ Rev.\ D {\bf 85}, 065008 (2012)
  [arXiv:1106.1167 [astro-ph.SR]].

\bibitem{Pejcha:2011en}
  O.~Pejcha, B.~Dasgupta and T.~A.~Thompson,
  ``Effect of Collective Neutrino Oscillations on the Neutrino Mechanism of Core-Collapse Supernovae,''
  Mon.\ Not.\ Roy.\ Astron.\ Soc.\  {\bf 425} (2012) 1083
  [arXiv:1106.5718 [astro-ph.HE]].
  
\bibitem{Duan:2010af} 
  H.~Duan, A.~Friedland, G.~C.~McLaughlin and R.~Surman,
  ``The influence of collective neutrino oscillations on a supernova r-process,''
  J.\ Phys.\ G {\bf 38}, 035201 (2011)
  [arXiv:1012.0532 [astro-ph.SR]].


\bibitem{Pllumbi:2014saa} 
  E.~Pllumbi, I.~Tamborra, S.~Wanajo, H.~T.~Janka and L.~Hüdepohl,
  ``Impact of neutrino flavor oscillations on the neutrino-driven wind nucleosynthesis of an electron-capture supernova,''
  Astrophys.\ J.\  {\bf 808}, no. 2, 188 (2015)
  [arXiv:1406.2596 [astro-ph.SR]].


\bibitem{Dasgupta:2011wg}
  B.~Dasgupta and J.~F.~Beacom,
  ``Reconstruction of supernova $\nu_\mu$, $\nu_\tau$, anti-$\nu_\mu$, and anti-$\nu_\tau$ neutrino spectra at scintillator detectors,''
  Phys.\ Rev.\ D {\bf 83} (2011) 113006
  [arXiv:1103.2768 [hep-ph]].
 
\bibitem{Serpico:2011ir} 
  P.~D.~Serpico, S.~Chakraborty, T.~Fischer, L.~Hudepohl, H.~T.~Janka and A.~Mirizzi,
  ``Probing the neutrino mass hierarchy with the rise time of a supernova burst,''
  Phys.\ Rev.\ D {\bf 85}, 085031 (2012)
  [arXiv:1111.4483 [astro-ph.SR]].
 
\bibitem{Borriello:2012zc} 
  E.~Borriello, S.~Chakraborty, A.~Mirizzi, P.~D.~Serpico and I.~Tamborra,
  ``(Down-to-)Earth matter effect in supernova neutrinos,''
  Phys.\ Rev.\ D {\bf 86}, 083004 (2012)
  [arXiv:1207.5049 [hep-ph]].
 
\bibitem{Beacom:2003nk} 
  J.~F.~Beacom and M.~R.~Vagins,
  ``GADZOOKS! Anti-neutrino spectroscopy with large water Cherenkov detectors,''
  Phys.\ Rev.\ Lett.\  {\bf 93}, 171101 (2004)
  [hep-ph/0309300].
 
\bibitem{SuperK-Gd}
L.~M.~Magro, talk given at Particle Physics and Cosmology Workshop 2015, Center for Theoretical Underground Physics and Related Areas, Deadwood, June 30, 2015.
   
\bibitem{Lunardini:2012ne} 
  C.~Lunardini and I.~Tamborra,
  ``Diffuse supernova neutrinos: oscillation effects, stellar cooling and progenitor mass dependence,''
  JCAP {\bf 1207}, 012 (2012)
  [arXiv:1205.6292 [astro-ph.SR]].
   
\bibitem{Banerjee:2011fj}
  A.~Banerjee, A.~Dighe and G.~Raffelt,
  ``Linearized flavor-stability analysis of dense neutrino streams,''
  Phys.\ Rev.\ D {\bf 84} (2011) 053013
  [arXiv:1107.2308 [hep-ph]].

\bibitem{Matt}
L.~Wolfenstein,  ``Neutrino Oscillations In Matter,'' Phys.\ Rev.\ D {\bf 17}, 2369 (1978); S.P.~Mikheev and A.Yu.\ Smirnov,   ``Resonance Enhancement Of Oscillations In Matter And Solar Neutrino Spectroscopy,'' Yad.\ Fiz.\ {\bf 42}, 1441 (1985) [Sov.\ J.\ Nucl.\ Phys.\ {\bf 42}, 913 (1985)]. 
  
\bibitem{Raffelt:2007yz} 
  G.~G.~Raffelt and G.~Sigl,
  ``Self-induced decoherence in dense neutrino gases,''
  Phys.\ Rev.\ D {\bf 75}, 083002 (2007)
  [hep-ph/0701182].

  
\bibitem{EstebanPretel:2007ec} 
  A.~Esteban-Pretel, S.~Pastor, R.~Tom{\`a}s, G.~G.~Raffelt and G.~Sigl,
  ``Decoherence in supernova neutrino transformations suppressed by deleptonization,''
  Phys.\ Rev.\ D {\bf 76}, 125018 (2007)
  [arXiv:0706.2498 [astro-ph]].


\bibitem{Sawyer:2008zs} 
  R.~F.~Sawyer,
  ``The multi-angle instability in dense neutrino systems,''
  Phys.\ Rev.\ D {\bf 79}, 105003 (2009)
  [arXiv:0803.4319 [astro-ph]].
  
\bibitem{pvtcom}  
 S. Chakraborty, R. S. Hansen, I. Izaguirre, and G. Raffelt have pointed out 
 to us that the instability footprints always tend to have the ``nose'' like 
 feature on the log-log scales shown here. The schematic has been improved to 
 reflect this feature.
  
  \bibitem{spectral}
  J.~P.~Boyd, ``Chebyshev and Fourier Spectral Methods,''
Second edition, Dover, New York (2001).

\bibitem{Sawyer:2005jk} 
  R.~F.~Sawyer,
  ``Speed-up of neutrino transformations in a supernova environment,''
  Phys.\ Rev.\ D {\bf 72}, 045003 (2005)
  [hep-ph/0503013].
  
\bibitem{wip}
F.~Capozzi, B.~Dasgupta and A.~Mirizzi, work in progress.  

\bibitem{Chakraborty:2011nf} 
  S.~Chakraborty, T.~Fischer, A.~Mirizzi, N.~Saviano and R.~Tom{\`a}s,
  ``No collective neutrino flavor conversions during the supernova accretion phase,''
  Phys.\ Rev.\ Lett.\  {\bf 107}, 151101 (2011)
  [arXiv:1104.4031 [hep-ph]].
  
\bibitem{Chakraborty:2011gd} 
  S.~Chakraborty, T.~Fischer, A.~Mirizzi, N.~Saviano and R.~Tom{\`a}s,
  ``Analysis of matter suppression in collective neutrino oscillations during the supernova accretion phase,''
  Phys.\ Rev.\ D {\bf 84}, 025002 (2011)
  [arXiv:1105.1130 [hep-ph]].
  
\bibitem{Sarikas:2011am} 
  S.~Sarikas, G.~G.~Raffelt, L.~Hudepohl and H.~T.~Janka,
  ``Suppression of Self-Induced Flavor Conversion in the Supernova Accretion Phase,''
  Phys.\ Rev.\ Lett.\  {\bf 108}, 061101 (2012)
  [arXiv:1109.3601 [astro-ph.SR]].
  
\bibitem{Chakraborty:2014nma} 
  S.~Chakraborty, A.~Mirizzi, N.~Saviano and D.~d.~S.~Seixas,
  ``Suppression of the multi-azimuthal-angle instability in dense neutrino gas during supernova accretion phase,''
  Phys.\ Rev.\ D {\bf 89}, no. 9, 093001 (2014)
  [arXiv:1402.1767 [hep-ph]].
 
\bibitem{Chakraborty:2014lsa} 
  S.~Chakraborty, G.~Raffelt, H.~T.~Janka and B.~Mueller,
  ``Supernova deleptonization asymmetry: Impact on self-induced flavor conversion,''
  arXiv:1412.0670 [hep-ph].

\end{thebibliography}
\end{document}